\documentclass[reprint,floatfix, twocolumn]{revtex4-1}

\usepackage{pifont} 

\usepackage[utf8]{inputenc}
\usepackage[a4paper]{geometry}
\usepackage{hyperref}
\usepackage{graphicx}
\usepackage{xcolor}
\usepackage{multirow}
\usepackage[normalem]{ulem}

\usepackage{setspace}
\usepackage[caption=false]{subfig}
\usepackage[T1]{fontenc}

\usepackage{amsmath}
\usepackage{amsfonts, relsize, color}
\usepackage{mathrsfs}
\usepackage{mathtools}
\usepackage{physics}
\usepackage{bbm}

\newcommand{\ncmd}{\newcommand}
\ncmd{\nn}{\nonumber}
\ncmd{\mbf}[1]{\bs{#1}}
\ncmd{\gam}{\gamma}
\ncmd{\sig}{\sigma}
\ncmd{\pha}{\alpha}
\ncmd{\lam}{\lambda}
\ncmd{\kap}{\kappa}
\ncmd{\Lam}{\Lambda}
\ncmd{\Gam}{\Gamma}
\ncmd{\Ups}{\Upsilon}
\ncmd{\Om}{\Omega}
\ncmd{\eps}{\epsilon}
\ncmd{\veps}{\varepsilon}
\ncmd{\vphi}{\varphi}
\ncmd{\vtheta}{\vartheta}
\ncmd{\tw}{\text{w}}
\ncmd{\pd}{\partial}
\ncmd{\pll}{\parallel}
\ncmd{\mc}{\mathcal}
\ncmd{\mf}{\mathfrak}
\ncmd{\dl}{\delta}
\ncmd{\bs}{\vec}
\ncmd{\half}{\frac{1}{2}}
\ncmd{\tilJ}{\widetilde{J}}
\ncmd{\avg}[1]{\langle{#1}\rangle}
\ncmd{\note}[1]{{\color{red}{\ding{168} [#1]}}}
\ncmd{\eq}[1]{Eq. \eqref{#1}}
\ncmd{\fig}[1]{Fig. \ref{#1}}
\ncmd{\suppl}{\note{`Supplementary Information'}}

\ncmd{\sur}[1]{{\color{blue}#1}}

\begin{document}

\title{Correlated Hopf Insulators}

\author{Konstantinos Ladovrechis$^1$ and Shouvik Sur$^2$}
\affiliation{$^1$ 
	Institute for Theoretical Physics and W\"{u}rzburg-Dresden Cluster of Excellence ct.qmat,
	Technische Universit\"{a}t Dresden, 01069 Dresden, Germany}
\affiliation{$^2$Department of Physics and Astronomy, Extreme Quantum Materials Alliance,
Smalley-Curl Institute, Rice University, Houston, Texas 77005, USA}

\begin{abstract}
Hopf insulators represent an exceptional class of topological matter unanticipated by the periodic table of topological invariants. 
These systems point to the existence of previously unexplored states of matter with unconventional topology. 
In this work, we take a step toward exploring this direction by investigating correlation-driven instabilities of Hopf insulators. 
Organizing our analysis around the topological quantum critical point that separates the Hopf insulating phase from a trivial insulator, we demonstrate the emergence of unconventional Weyl semimetallic and topological insulating states. 
Notably, the Weyl semimetal supports non-reciprocal superconductivity and a Bogoliubov-Fermi surface, potentially providing a novel framework for realizing the superconducting diode effect. 
Finally, we highlight the interconnectedness of the effective descriptions of correlated Hopf insulators, two-dimensional quadratic band-touching semimetals, and Luttinger semimetals.
\end{abstract}

\date{\today}

\maketitle

\section{Introduction}
In topological materials, interaction-driven phase transitions either leave an imprint of the topology of the parent state on the daughter states~\cite{pesin2010mott, li2018topological, setty2024symmetry, setty2023topological}, or generate new topological features~\cite{raghu2008topological,Dzero2010,sun2009,Alicea2012, Beenakker2013,Volovik2019,lu2022interaction}.
In this sense, as a driver of topological phase transitions, inter-particle interactions provide a key mechanism  for tuning between different topological phases of matter. 
While across a topological phase transition the global topology of the respective ground states changes, the topology in  lower dimensional subspaces of the Brillouin zone may remain intact.
This is particularly true for symmetry-protected topological phases which are protected by a combination of space-group symmetries. 
Topological features protected by space-group symmetries that are preserved across a symmetry-breaking phase transition, in principle, can survive in the symmetry-broken state.
Hopf insulators formally lie beyond the tenfold way~\cite{moore2008}. They  possess unconventional topological features characterized by delicate and multi-cellular topologies~\cite{nelson2021a} which can be diagnosed by the returning Thouless pump (RTP)~\cite{nelson2022} and the staggered Chern number~\cite{tyner2024dipolar}. 
Over the past decade and a half, various properties of these enigmatic insulators have been explored, both theoretically and on engineered platforms~\cite{deng2013Hopf, kennedy2016,liu2017,deng2018,schuster2019a,schuster2019b,schuster2021,yuan2017,unal2019,wang2023}. In recent years, there has been a resurgence of interest in topological states sharing key features with Hopf insulators, leading to proposals for new  types of band-topological states~\cite{zhu2023z2,ezawa2017, tyner2024dipolar,graf2023, parasar2024delicate, zhuang2024berry}, and advancement in our understanding of the topology in more conventional states~\cite{zhu2023Delicate}.
Thus, it is pertinent to inquire whether correlation effects can help generate hitherto unexplored states which inherit topological features from the parent Hopf insulating state.

In this work, we employ strong short-range interactions to destabilize  fourfold rotation symmetry-protected Hopf insulators, leading to the emergence of novel insulating and semimetallic states. 
Our aim is twofold: first, to examine the stability of Hopf insulators in the presence of strong short-range interactions, and second, to explore whether previously unrecognized topologically non-trivial states of matter exist that share the similar  topological obstructions as Hopf insulators.
To maintain analytical control over the interplay between band topology and electronic correlations, we anchor our analysis to  the topological quantum critical point (TQCP)  which separates a Hopf insulator from a trivial insulator.
At the TQCP a quadratic band-touching point (QBT) is realized at a high-symmetry point in the Brillouin zone, and it exemplifies  an emergent Berry dipole~\cite{nelson2022}. 
We demonstrate that this TQCP is the three-dimensional analog of the two-dimensional QBT semimetals found in systems such as bilayer graphene~\cite{mccann2006landau, castro2009electronic} and the checkerboard lattice~\cite{sun2009, sur2018}. 
Additionally, the TQCP shares similarities with the band-touching point in Luttinger semimetals~\cite{luttinger1956,zhu2023Delicate}, which occur in various families of materials, including $\alpha$-tin~\cite{groves1963band,roman1972stress}, pyrochlore iridates~\cite{yang2010topological,wan2011, witczak2014correlated,kondo2015quadratic}, and half-Heusler compounds~\cite{lin2010half,liu2016observation}. 
Thus, a question of general interest is to understand the extent to which the physics of correlated Hopf insulators in the vicinity of the TQCP resemble those in the more conventional systems.

\begin{figure}[!t]
\centering
\includegraphics[width=0.9\columnwidth]{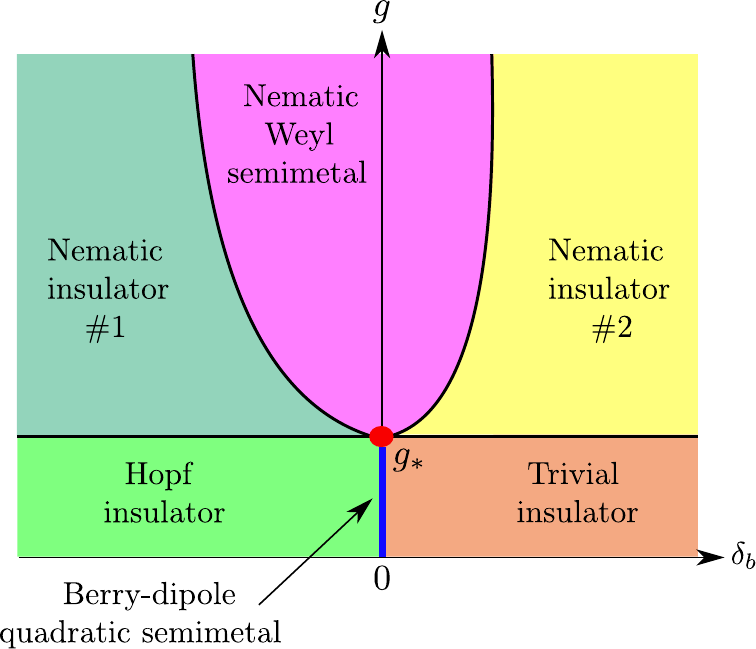}
\caption{Schematic phase diagram of a correlated Hopf insulator. 
The sign of the topological mass parameter, $\delta_b$, distinguishes the two non-interacting phases around the topological quantum critical point (TQCP) at  $\delta_b=0$, viz.,  Hopf insulator $(\delta_b<0)$ and topologically trivial insulator $(\delta_b>0)$. 
The parameter $g$ represents the strength of the on-site repulsive Hubbard interaction (see Eq.~\eqref{eq:FierzBasis}). 
The red dot is a multi-critical point which is deduced from the interacting unstable fixed point obtained in our renormalization group (RG) analysis (see Fig.~\ref{fig:WilsonFisher}).
The phase boundary separating the rotation-symmetric and nematic insulators (nematic insulators and the Weyl semimetal) is deduced from the separatrices in   Fig.~\ref{fig:WilsonFisher} (the Chern number supported by the $k_z = 0$ plane of the Brillouin zone as shown in Fig.~\ref{fig:chern-num}).}
\label{fig:phase-diag}
\end{figure}

By a combination of renormalization group (RG) and topological diagnostics we show that while the insulating phases and the semimetal at the TQCP, obtained in the non-interacting limit, are perturbatively stable against interactions, sufficiently strong interactions drive a nematic instability, characterized by a spontaneous breaking of the fourfold rotational symmetry (see Fig.~\ref{fig:phase-diag}). 
In particular, if the repulsive interaction strength $g$ exceeds a critical value, $g_*$, then the semimetallic state at the TQCP becomes unstable to a Weyl semimetal (WSM) state. 
The Weyl points, thus obtained, may be considered as nematicity-induced deformation of the emergent Berry dipole.
Due to a lack of time-reversal, inversion, and mirror symmetries, the WSM state exhibits highly anisotropic distribution of Weyl points, and it supports Fermi arcs on surface terminations perpendicular to the principal axes.
Away from the TQCP, the WSM state is separated from the Hopf and  trivial insulating phases by four-fold rotational symmetry-broken insulating states -- ``Nematic insulator $\#1$'' and ``Nematic insulator $\#2$'', as depicted in Fig.~\ref{fig:phase-diag}.
While no clear topological distinction exists between ``Nematic insulator $\#2$'' and the trivial insulator, the RTP invariant in ``Nematic insulator $\#1$'' is halved compared to that in the Hopf insulator.
Importantly, we show that all three nematic states can support non-reciprocal superconductivity upon doping, providing a potential pathway for realizing the superconducting diode effect~\cite{Nadeem2023,Ando2020,Wu2022,Lin2022,davydova2024}.

The paper is organized as follows. In Section~\ref{sec:Model} we introduce the model of four-fold rotationally symmetric interacting Hopf insulators, and we discuss its connections to other conventional two and three dimensional phases. 
In Section~\ref{sec:RG} we present results from our RG analysis in the vicinity of the TQCP, and we  identify the dominant instabilities.
Section~\ref{sec:MeanField} is primarily dedicated to a mean-field level analysis of the phase diagram in Fig.~\ref{fig:phase-diag}, where (i) we obtain the phase boundary between the WSM and nematic insulating phases; 
(ii) we discuss the topological features of the symmetry-broken states, along with surface states.
In Section~\ref{sec:SC} we  briefly discuss how non-reciprocal superconductivity may be realized in this system. 
We conclude with a summary of our  results and an outlook in Section~\ref{sec:conclusion}.

\section{Model}\label{sec:Model}
We consider the Moore-Ran-Wen model of Hopf insulators specified by the Bloch Hamiltonian~\cite{moore2008}  
\begin{align}
H_0(\vec{k}, m) = \vec w(\vec{k}, m) \cdot \vec \sigma.
\label{eq:HopfLattice}
\end{align}
The three-dimensional vector $\vec w(\vec k, m)$ reads
\begin{equation}\label{eq:ColumnVector}
\vec w(\vec k, m) = \begin{bmatrix}
2u_1(\vec k)\,u_3(\vec k) + 2u_2(\vec k)\,u_4(\vec k, m)\\
2u_1(\vec k)\,u_4(\vec k, m) - 2u_2(\vec k)\,u_3(\vec k)\\
u_1^2(\vec k) + u_2^2(\vec k) - u_3^2(\vec k) - u_4^2(\vec k, m)
\end{bmatrix},
\end{equation}
where $u_1(\vec k)=t \sin{k_x}$, $u_2(\vec k)=t \sin{k_y}$, $u_3(\vec k)=t \sin{k_z}$, and $u_4(\vec k, m)=t_0 \qty[\cos{k_x}+ \cos{k_y}+ \cos{k_z} - m]$ with the parameters $t$ and $t_0$ controlling the bandwidth. 
Henceforth, we measure energy in units of $t_0$, and set $t_0 = 1$.
The Pauli matrices act on an internal degree of freedom, such as a sublattice index, $A$ and $B$. 
The four-fold rotation symmetry about the $\hat k_z$ axis manifests itself through the operation
\begin{flalign}\label{eq:C4z}
    \mc C_{4}^z:\quad H_0(\vec k,m)\,\, \mapsto\,\, &H_0(k_y, -k_x, k_z,m)\nonumber\\
    &= e^{i \frac{\pi}{4}\sigma_3} H_0(\vec k,m) e^{-i \frac{\pi}{4}\sigma_3}.
\end{flalign}
Eq.~\eqref{eq:HopfLattice} also respects the anti-unitary mirror operations~\cite{tyner2024dipolar} (suppressing reference to $m$)
\begin{flalign}\label{eq:MxMy}
    &\mc M_x \mc T:\quad H_0(\bs k)\,\, \mapsto\,\, H_0^*(k_x, -k_y, -k_z) = \sig_3 H_0(\bs k) \sig_3,\nonumber\\
    &\mc M_y \mc T:\quad H_0(\bs k)\,\, \mapsto\,\, H_0^*(-k_x, k_y, -k_z) = H_0(\bs k),
\end{flalign}
where $\mc T$ implements spinless time-reversal.
For $|m|<3$, $H_0(\bs k, m)$ describes three distinct Hopf-insulating phases, separated by two TQCPs at $m = \pm  1$~\cite{moore2008,deng2013Hopf}. 
The TQCPs at $m=\pm 3$ separate the Hopf insulator with Hopf invariant $-1$ from ordinary insulators. 
Here, we will focus on the TQCP at $m = 3$, where the two  bands, resulting from diagonalizing $H_0(\vec k,m)$, touch quadratically at the center of the Brillouin zone, i.e.~the $\Gam$ point. 
This band-touching point acts as a simultaneous source and  sink of Berry curvature, and it  asymptotically exhibits a quantized Berry-dipole moment~\cite{nelson2022}. 
We note that the $\mc C_{4}^z$ symmetry  allows a topological classification of the eigenstates of $H_0(\vec k,m)$ by the returning Thouless pump, which can be related to the Hopf invariant~\cite{nelson2021a,nelson2022}. 
In the rest of the paper, we will treat the deviation from the TQCP at $m=3$
as a \textit{topological mass} that tunes the system between Hopf and ordinary insulating phases. 
Due to the presence of the semimetallic state at $m = 3$, the TQCP provides a convenient reference point for a scaling analysis which we develop and apply to investigate the interacting phase diagram of the Hopf insulator in the vicinity of the TQCP.

The fermionic field operator is defined as the two-dimensional spinor
$\hat\Psi^\dag(\vec{R},\tau)=\left[\hat{c}^\dag_A(\vec{R},\tau),\,\hat{c}^\dag_B(\vec{R},\tau)\right]$
where the operator $\hat{c}^\dag_{A\,(B)}(\vec{R},\tau)$ creates an electron at site $A$ ($B$) in the unit cell located at position $\vec R$ at Euclidean time $\tau$.
The most general local Lagrangian term representing short-range interactions  would be of the form $L_\text{int} =\sum_{\vec R}[\hat\Psi^\dag(\vec{R},\tau)M\hat\Psi(\vec{R},\tau)][\hat\Psi^\dag(\vec{R},\tau)N\hat\Psi(\vec{R},\tau)]$, where $M$ and $N$ are $2\times2$ Hermitian matrices. 
Although 10 independent terms are present in $L_\text{int}$, the $C^z_4$ symmetry will reduce the number as follows.
The four possible $2\times2$ Hermitian matrices, $\{\mathbbm{1}_2, \sigma_{j=1,2,3}\}$,  are classified into two categories according to whether they transform as \textit{scalars}, $S=\{\mathbbm{1}_2,\,\sigma_3\}$, or \textit{vectors}, $V=\{\sigma_1,\,\sigma_2\}$, under $C^z_4$.
Therefore, the $C^z_4$-invariant interactions contain four independent coupling constants,
\begin{flalign}
L_\text{int,$C_4$}=\sum^2_{i\geq j=1}s_{ij}\,[\hat\Psi^\dag(\vec{R},\tau)S_i\hat\Psi(\vec{R},\tau)][\hat\Psi^\dag(\vec{R},\tau)S_j\hat\Psi(\vec{R},\tau)]\nn\\ +v\left([\hat\Psi^\dag(\vec{R},\tau)\sigma_1\hat\Psi(\vec{R},\tau)]^2 +[\hat\Psi^\dag(\vec{R},\tau)\sigma_2\hat\Psi(\vec{R},\tau)]^2\right)
\end{flalign}
with $s_{ij},\,v\geq0$. 
This number is further reduced due to the fermionic Fierz transformation \cite{fierz1937,nishi2005,herbut2009}. 
The latter imposes the following linear constraints on the four coupling constants, $v=2s_{22}=-2s_{11}$ and $s_{21}=0$. 
Therefore, there is only one linearly independent channel forming the Fierz-complete basis for the interacting Lagrangian. 
Choosing this channel to correspond to the  matrix $\mathbbm{1}_2$, we write $(g\geq0)$
\begin{align}\label{eq:FierzBasis}
L_{\text{int}}&= g \sum_{\vec R}\left[\hat{\Psi}^\dag(\vec R,\tau)\,\hat{\Psi}(\vec R,\tau)\right]^2.
\end{align}
Since the model describes interacting spinless electrons, unsurprisingly this interaction vertex is equivalent to the on-site repulsive Hubbard interaction, $\sum_{\vec R} \hat{n}_A(\vec R,\tau)\,\hat{n}_B(\vec R,\tau)$, 
where $\hat{n}_i(\vec R,\tau)$
is the density operator for the $i$-th sublattice with $i \in \{A,B\}$.

In general, interactions among degrees of freedom can drive spontaneous symmetry breaking, which  raises the question on the stability of topological insulators and semimetals in the presence of inter-particle interactions. 
With the exception of two-dimensional quadratic band-touching semimetals~\cite{sun2009, vafek2010}, topological insulators and semimetals in two or more spatial dimensions are  robust against arbitrarily weak short-range interactions~\cite{herbut2009theory, maciejko2014weyl, sur2016instabilities, roy2017interacting}. 
The robustness of the latter category of states can be attributed to a vanishing low-energy density of states which effectively weakens the impact of interactions.
Therefore, any potential instability must arise from interactions that are sufficiently strong to overcome the suppression due to the vanishing density of states.
In Secs.~\ref{sec:RG} and \ref{sec:MeanField} we will discuss the patterns of symmetry breaking. We conclude the present section with a description of the TQCP in the non-interacting limit which will anchor the analyses to follow.

\subsection{Interacting Berry dipole}\label{sec:dipole}
In the vicinity of the TQCP the low-energy physics is predominantly controlled by momentum states close to the $\Gamma$ point. 
In this regard, we expand  $H_0(\vec k,m)$ to quadratic order in the deviation from the $\Gamma$ point. 
We supplement the resultant $k.p$ model with the Fierz-complete interaction basis given in Eq.~\eqref{eq:FierzBasis}
to obtain the effective action 
\begin{align}
S(\delta_b) =& \int \dd{k} \Psi^\dagger(k)\qty[ik_0 \mathbbm{1}_2 - h_0(\vec k, \delta_b)]\Psi(k)\nn\\
&+\dfrac{g}{2} \int\dd{k} n_A(-k) n_B(k).
\label{eq:action}
\end{align}
where $k_0$ is the Euclidean frequency, the integration measure is $\dd{k} \equiv \frac{dk_0 d^3 k}{(2\pi)^4}$, $\delta_b = m - 3$ parameterizes the topological mass, 
and the effective single-particle Hamiltonian 
\begin{align}\label{eq:QuadraticHamiltonian}
h_0(\vec k,\delta_b) =&
2\qty[t^2 k_x k_z +t  k_y \delta_b]\sigma_1+2 \qty[t^2 k_y k_z - t k_x \delta_b]\sigma_2 \nn \\
&\qquad + \qty[t^2(k_x^2+ k_y^2-k^2_z)- \delta_b^2]\sigma_3.
\end{align}
We note that the  topological mass takes an unusual form in Eq.~\eqref{eq:QuadraticHamiltonian}, $\qty[2t k_y \sigma_1 - 2t k_x \sigma_2] \delta_b-\delta_b^2\sigma_3$, as it is spanned by all three Pauli matrices. 
Through the (inverse) Hopf map it is related to the topological mass in centrosymmetric three-dimensional strong topological insulators that gaps the Dirac point at the center of the Brillouin zone~\cite{murakami2007phase}.
Because of ``linearization'' of the dispersion in the vicinity of a high-symmetry point in the Brillouin zone, new \textit{effective symmetries} may emerge.
Here, a $z$-mirror symmetry ($\mc M_z$) emerges at the TQCP which is represented by the operation 
\begin{flalign}\label{eq:ZMirrorSymmetry}
    \mathcal{M}_z:\quad h_0(\vec k, \delta_b=0)\,\, \mapsto\,\, &h_0(k_x, k_y, -k_z,\delta_b=0)\nonumber\\
    &= \sigma_3 h_0(\vec k, \delta_b = 0) \sigma_3,
\end{flalign}    
Therefore, both a finite $\dl_b$ and  the inclusion of $\order{|\vec k|^3}$ terms in $h_0(\vec k,\delta_b)$
will also break $\mathcal{M}_z$.
We note that while the former is  a relevant perturbation within the scaling theory discussed below, the latter is irrelevant.

The emergent $z$-mirror symmetry plays a vital role in ensuring that the Berry curvature emanating from the QBT at $(\vec k,\delta_b)=(\vec{0},0)$ supports a quantized dipolar flux~\cite{nelson2022}.  
In particular, the radial component of the Berry curvature on a  Gaussian sphere enclosing the QBT is given by~\cite{tyner2024dipolar}
\begin{align}\label{eq:BerryRadialComponent}
B_r(|\vec k|, \theta, \phi) = \frac{\cos\theta}{\pi |\vec k|^2},
\end{align}
where we have employed the spherical coordinate system in the momentum space such that $(k_x, k_y, k_z) = |\vec k|(\sin\theta \cos\phi, \sin\theta \sin\phi, \cos\theta)$.
The vanishing of $ \int_{0}^{2\pi} \int_0^\pi \dd{\phi} \dd{\theta} \sin\theta ~ B_r$ indicates the net Berry flux passing through the Gaussian sphere is zero and, thus,  the QBT is not a Berry monopole.
By contrast, the staggered-Berry flux through the Gaussian sphere,  $\mf C_{\text{stagg}} = |\vec k|^2 \int_{0}^{2\pi} \int_0^\pi \dd{\phi} \dd{\theta} \sin\theta 
~\mbox{sign}(\pi/2-\theta)~ B_r$, equals $2$ in units of $2\pi$. 
It implies the northern and southern hemispheres of the Gaussian sphere enclose opposite Berry fluxes of magnitude $2\pi$.
In this sense the QBT is a Berry dipole.

\subsection{Connections to other models}
\begin{figure}[!t]
\centering
\includegraphics[width=0.7\columnwidth]{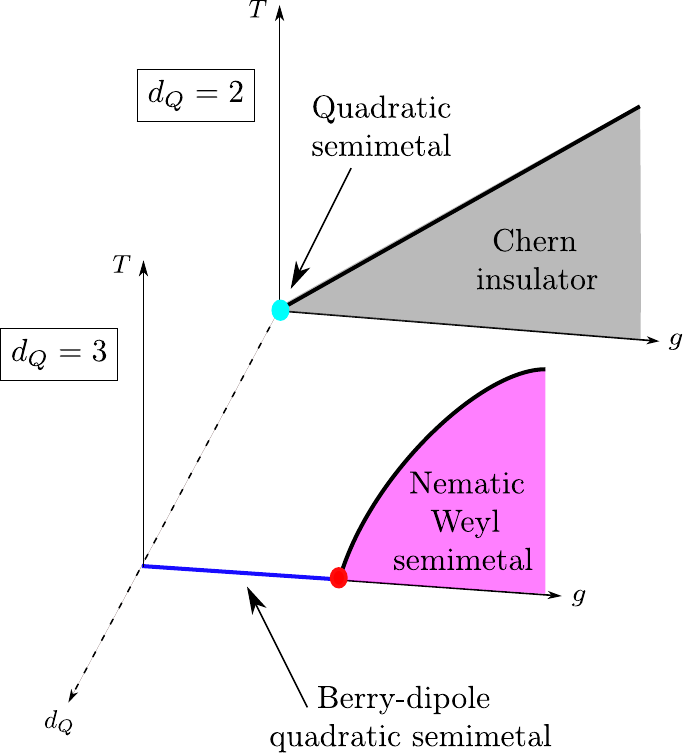}
\caption{Interacting phase diagram connecting the candidate phases in $d=2$ and $d=3$ spatial dimensions. 
Since we focus on correlated quadratic band-touching semimetals (QBTSM), the number of quadratically dispersing directions in the non-interacting limit, $d_Q = d$. 
At $d_Q=2$ ($d_Q=3$), the QBTSM is perturbatively unstable (stable)  against interactions.
Consequenly, arbitrarily weak (strong) interactions drive the system to a Chern insulating (nematic Weyl semimetallic) state. 
In Sec.~\ref{sec:MeanField} we show how the Chern insulating layer is embedded within the Weyl semimetal.}
\label{fig:inter-phase-diag}
\end{figure}

The QBT semimetal described by $h_0(\bs k, 0)$ is connected to other well-known semimetallic states with QBT both in two and three (spatial) dimensions.
To demonstrate that correspondence, we generalize Eq.~\eqref{eq:QuadraticHamiltonian} at $\delta_b=0$ to $d$ spatial dimensions with $d$ $\in$ $[2,3]$ as  (see also Appendix \ref{sec:RenormalizationGroup})
\begin{align}
    h_0(\bs k, 0) \to  h'_{0,d}(\bs k) = 2 t^2 k_d \bs k_\perp \cdot \bs \gamma + t^2(k_d^2 - k_\perp^2) \gamma_d,
\label{eq:general-h}
\end{align}
where $\bs k_\perp= (k_1, \ldots, k_{d-1})$ and $\{\gamma_1,\dots,\gamma_d\}$ is a set of mutually anticommuting matrices. 
We note that $h'_{0,d}(\bs k)$ is rotationally symmetric in the $\bs k_\perp$-subspace .  
While the TQCP for the Hopf insulator corresponds to the case $d = 3$, the two-dimensional QBT semimetal arising in, for example, Bernal-stacked bilayer graphene and the checkerboard lattice, corresponds to $d=2$.
Therefore,  $h_0(\bs k, 0) \equiv h_{0,3}'(\bs k) $ is connected to the two-dimensional QBT semimetals, $h_{0,2}'(\bs k,0)$, via dimensional tuning, whereby we continuously tune the number of  components of $\bs k_\perp$~\cite{zinn2021quantum}.
In Sec.~\ref{sec:RG}, we will use this connection to perform a controlled renormalization-group analysis by tuning the number of quadratically dispersing directions, $d_Q$~\cite{sur2019}. 
We will also demonstrate that $d_Q$ as a tuning parameter is not only a formal tool for controlling quantum fluctuations, but it also connects  the phase diagram at $d_Q = 2$~\cite{sun2009, sur2018} to that obtained here (cf. Fig.~\ref{fig:phase-diag}), as illustrated  in Fig.~\ref{fig:inter-phase-diag}.

A different generalization of
$h_0(\bs k,0)$ is possible by fixing $d_Q = 3$ but increasing the dimension of representation of $\gamma_n$ from two to four. 
This connects the TQCP of the Hopf insulator to Luttinger semimetals, which is described by the effective  Bloch Hamiltonian, 
\begin{align}
h''_{0,3}(\bs k) =& t' k_z \bs k_\perp \cdot \bs \gamma + t'(2k_z^2 - k_\perp^2) \gamma_{3} + t'' k_x k_y \gamma_4 \nn \\ 
& \quad + t''(k_x^2 - k_y^2) \gamma_5.
\end{align}
The QBT obtained by setting $t'' = 0$ can be thought of as being composed of two copies of oppositely charged Berry dipoles, in analogy to certain models of Dirac semimetals in which a Dirac point corresponds to two copies of oppositely charged Weyl points.

\section{Renormalization group}\label{sec:RG}
Having described the non-interacting model in Eq.~\eqref{eq:QuadraticHamiltonian}, valid in the vicinity of a TQCP separating a Hopf from a topologically trivial insulator, in the current section we present the renormalization-group analysis which sets the context for the correlated phase diagram given in Fig.~\ref{fig:phase-diag}.
{In the interest of performing dimensional regularization in the interacting action given by Eq.~\eqref{eq:action}, we describe the scaling behavior in generic $d$ spatial dimensions following the generalization scheme described in Eq.~\eqref{eq:general-h} and Appendix \ref{sec:RenormalizationGroup}.}
The non-interacting part, $h_{0,d}(\vec{k},\delta_b)$, is invariant under spatiotemporal scaling, which leads to the following engineering dimensions: $[k_j] = L^{-1}$, $[k_0] = L^{-2}$, $[t] = L^0$, and $[\Psi] = L^{(4 + d)/2}$ with $L$ being a length scale.
Here, $t$ is treated as a dimensionless parameter, and the bare dynamical critical exponent is $z = 2$.
Therefore, the parameter controlling the topological mass, $\delta_b$, has engineering dimension $[\delta_b]=L^{-1}$, {while the coupling constant} $g$ scales as $[g] = L^{d-2}$.
That implies that the bare topological mass is a relevant perturbation in the vicinity of the TQCP, and its renormalized value has to be fixed at zero in order to maintain the topological quantum criticality.
By contrast, $g$ is a relevant perturbation only for $d < 2$, which implies that the TQCP is perturbatively stable against short-ranged interactions above $d = d_c =2$, in analogy to Luttinger semimetals. 
{However}, it can be destabilized by a sufficiently strong interaction.
In the rest of this section we investigate the combined impact of $g$  and $\delta_b$ on the TQCP, and {we} obtain the phase diagram in Fig.~\ref{fig:phase-diag}.

\begin{figure}
\centering
\includegraphics[scale=0.5]{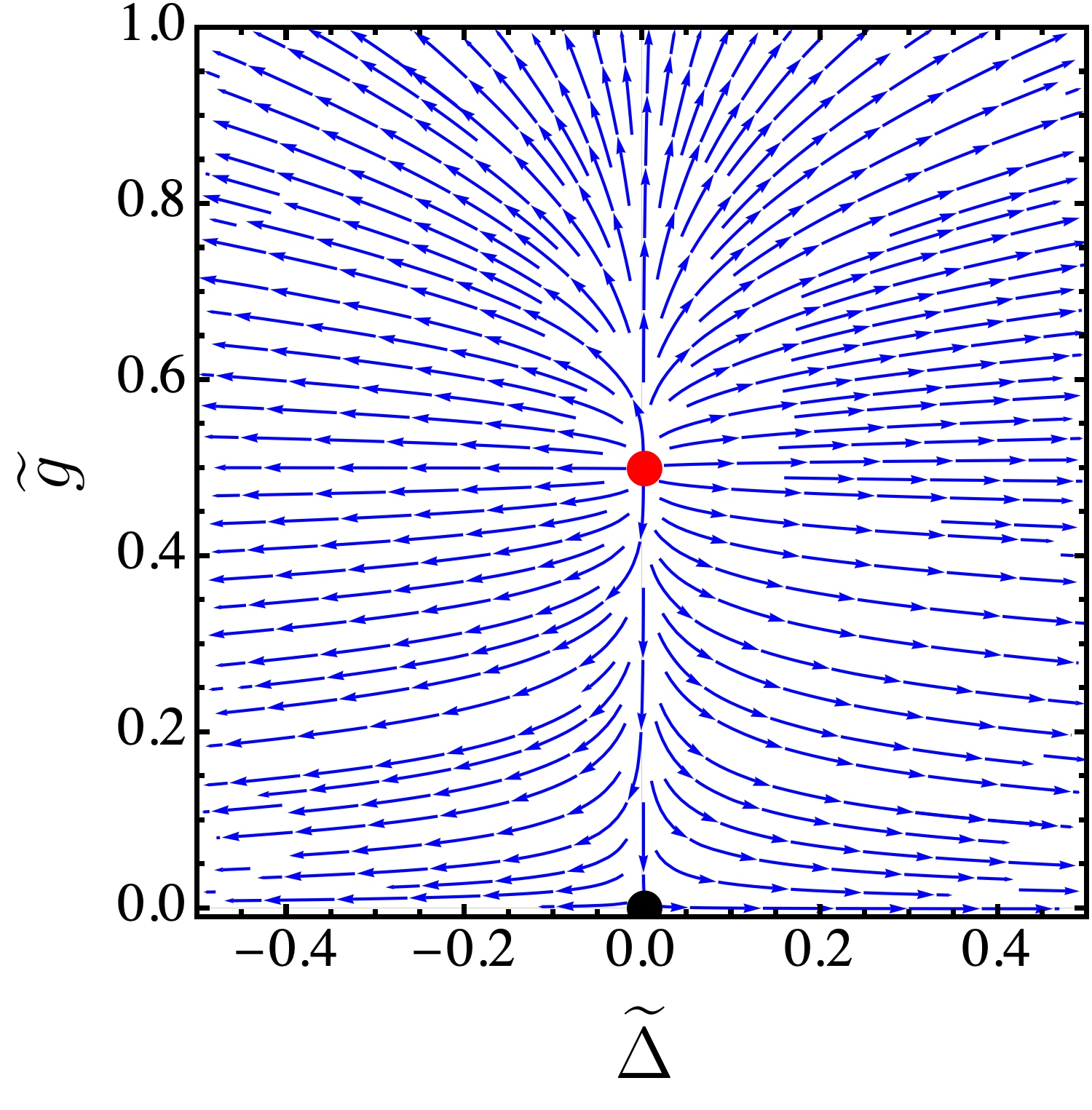}
\caption{Phase portrait in the phase space of the variables $(\widetilde\Delta,\widetilde{\textsl{g}})$ based on  Eqs.~\eqref{eq:TopologicalMassRGEquation} and \eqref{eq:BetaFunction} in the isotropic limit $t=1$ when $\epsilon=1$. The black dot denotes the Gaussian fixed-point solution at $(\widetilde\Delta_G,\widetilde{g}_G)$, and the red dot denotes the fixed-point solution $(\widetilde\Delta_\star,\widetilde{g}_\star)$ (see Eq.~\eqref{eq:FixedPointSolutions}).
The renormalization group flow implies that the former (latter) is a quantum-critical (bi-critical) point. 
Such a fixed-point structure resembles the behavior of the effective $\phi^4$ theory for the classical Ising model in $d>4$ spatial dimensions \cite{herbut2007,goldenfeld2018}.}
	\label{fig:WilsonFisher}
\end{figure}

Since the non-trivial impact of $g$ in $d=3$ only appears at strong coupling, we perform a Wilsonian renormalization-group analysis in $d =  2 + \eps = d_Q$ dimensions, {where the real-valued parameter $\epsilon$ acts as an \textit{expansion parameter} exerting control over the generated} quantum corrections.
Denoting the momentum UV cutoff by $\Lambda$, we use the radial component of the momentum vector $|\vec{k}|$  for defining the successive RG momentum shells as $\Lambda/\ell<|\vec{k}|<\Lambda$, with $1<\ell<\infty$ being the dimensionless scaling parameter.
The angular part of the  Feynman diagrams is calculated at $d=3$. Upon introducing the dimensionless quantities $\widetilde g=g(\Lambda/\ell)^{d-2}/(8\pi^2)$, $\widetilde\Psi=\Psi(\Lambda/\ell)^{-(4+d)/2}$, $\widetilde k=k(\Lambda/\ell)^{-1}$,
$\widetilde k_0=k_0(\Lambda/\ell)^{-2}$,
and $\widetilde{\delta}_b=\delta_b(\Lambda/\ell)^{-1}$, the one-loop renormalized action becomes self-similar to the original one given by Eq.~\eqref{eq:action} in the presence of the following two beta functions (see Appendix \ref{sec:RenormalizationGroup})
\begin{align}
& \beta_{\widetilde\Delta}:=\dfrac{d\widetilde\Delta}{d\text{ln}\ell} = \widetilde\Delta\left[2+\dfrac{8\widetilde{g}}{3t^2}\right], \label{eq:TopologicalMassRGEquation} \\
& \beta_{\widetilde g}:=\dfrac{d\widetilde g}{d\text{ln}\ell} = -\epsilon\widetilde g+\dfrac{2\widetilde{g}^2}{t^2}. \label{eq:BetaFunction}
\end{align}
Due to the nonvanishing Hartee-Fock self-energy contribution, we have defined the \textit{renormalized} {parameter} $\widetilde\Delta \coloneqq \widetilde\delta^2_b-\widetilde \Delta_\text{TQCP}$, where $\widetilde\Delta_\text{TQCP}=\widetilde g(1-\ell^{-2})/(24\pi^2)$  is the shifted TQCP location. Eqs.~\eqref{eq:TopologicalMassRGEquation} and \eqref{eq:BetaFunction} exhibit two fixed-point solutions:
\begin{equation}\label{eq:FixedPointSolutions}
(\widetilde\Delta_G,\widetilde{g}_G)=(0,0)\qquad\text{and}\qquad(\widetilde\Delta_\star,\widetilde{g}_{\star})=\left(0,\dfrac{t^2\epsilon}{2}\right).
\end{equation}
In the vicinity of each fixed-point solution, we use the stability matrix 
$B_{ij}=\partial \beta_i/\partial j$, where $i,j=\{\widetilde\Delta,\widetilde g\}$,
to determine that the Gaussian fixed point $(\widetilde\Delta_G,\widetilde{g}_G)$ is a \textit{critical point} with unstable direction along the $\widetilde\Delta$ axis, and the fixed-point solution $(\widetilde\Delta_\star,\widetilde{g}_{\star})$ is a \textit{bi-critical point} with unstable directions along both axes. 
The complete phase portrait of Eqs.~\eqref{eq:BetaFunction} and \eqref{eq:TopologicalMassRGEquation} has been numerically evaluated for $t=\epsilon=1$, and it is depicted in Fig.~\ref{fig:WilsonFisher}.
Due to the emergent $z$-mirror symmetry (see Eq.~\ref{eq:ZMirrorSymmetry}), the phase portrait is symmetric with respect to the line $\widetilde{\Delta}=0$, and it is qualitatively similar to the well-known Wilson-Fisher fixed-point structure of the bosonic $\phi^4$ theory describing the thermal phase transition of the classical Ising model in $d>4$ spatial dimensions \cite{herbut2007,goldenfeld2018}. 
For an initial value of the coupling constant $\widetilde g<\widetilde g_\star$, successive RG transformations take the systems towards a noninteracting phase, which can be a Hopf insulator $(\widetilde\Delta<0)$, a trivial insulator $(\widetilde\Delta>0)$, or the QBT semimetal $(\widetilde\Delta=0)$ that is present at the single-particle TQCP.

By contrast, for any initial value  of $\widetilde{g}>\widetilde{g}_\star$, the system is driven towards a strong-coupling instability.  
To identify the nature of the dominant strong-coupling instability, we perform a susceptibility analysis. To that end, we introduce to Eq.~\eqref{eq:action} the source term $M_j\bar\Psi(\vec{k},i\omega_n)\gamma_j\Psi(\vec{k},i\omega_n)$, where $j=\{0,1,2,3\}$ is an enumeration index with $\vec{\gamma}=\{\mathbbm{1}_2,\sigma_1,\sigma_2,\sigma_3\}$ and positive-valued parameters $\vec{M}=\{M_0,M_1,M_2,M_3\}$. 
We then evaluate the impact of the local interaction term (see Eq.~\eqref{eq:FierzBasis}) on the amplitude of each order parameter to one-loop order under successive RG transformations in three spatial dimensions. 
This leads to the following RG equations (see Appendix \ref{sec:RenormalizationGroup})
\begin{flalign}\label{eq:SusceptibilitiesRGEquations}
&\dfrac{d\text{ln}\widetilde M_0}{d\text{ln}\ell} = 2,\qquad\dfrac{d\text{ln}\widetilde M_1}{d\text{ln}\ell} 
= 2+\dfrac{22\widetilde{g}}{15t^2},\nonumber\\
&\dfrac{d\text{ln}\widetilde M_2}{d\text{ln}\ell} = 2+\dfrac{22\widetilde{g}}{15t^2},\qquad\dfrac{d\text{ln}\widetilde M_3}{d\text{ln}\ell} = 2+\dfrac{16\widetilde{g}}{15t^2},
\end{flalign}
where $\widetilde{M}_j=M_j\left(\Lambda/\ell\right)^{-2}$ is the corresponding dimensionless entity. 
Thus, the channels with parameters $M_1$ and $M_2$ are degenerate, and they constitute the dominant instabilities at strong coupling.
The corresponding symmetry-broken state is characterized by an arbitrary mixture of the two mass terms, $\cos\theta \langle\bar\Psi\sigma_x\Psi\rangle + \sin\theta \langle\bar\Psi\sigma_y\Psi\rangle$ with $\theta \in [0, 2\pi]$.
This degeneracy is a consequence of the {four-fold} rotational symmetry about the $\hat k_z$-axis in the parent state (i.e. on the $k_x - k_y$ plane of the three-dimensional system; cf. Eq.~\eqref{eq:HopfLattice}).
In the two-dimensional limit, this degeneracy is lost, and the  nucleation of  $\langle\bar\Psi\sigma_y\Psi\rangle$ drives the dominant instability to an anomalous Hall insulating state~\cite{sun2009}. 
{The form of the dominant instability in our case suggests that the four-fold rotational symmetry about the $\hat z$ axis is spontaneously broken in the strongly interacting regime, leading to a certain form of spatial anisotropy. Thus, the correlation-driven phases possess a \textit{nematic character}.}
We note that the anomalous dimensions of $\widetilde M_j$'s are 
independent of the {parameter} $\widetilde\Delta$ (or the topological mass $\delta_b$). Thus, the present susceptibility analysis holds in the vicinity of the TQCP and away from the limit $\delta_b=0$, as well.  

\section{Phase Diagram and new phases}\label{sec:MeanField}
\subsection{States at strong coupling}
Due to the bulk gap in the insulating phases supported by $H_0(\bs k, m)$ and the irrelevance of short-ranged interactions at the TQCP, only two phases, Hopf and ordinary insulators, are possible at weak interactions.
The preceding analyses revealed the existence of rotational symmetry breaking states at strong couplings. 
Here, we will discuss the nature of the new states that are stabilized by strong interactions. 
For this purpose, it will be convenient to express the mean-field Hamiltonian in the symmetry broken state as 
\begin{align}\label{eq:MeanFieldLattice}
H_{\text{MF}}(\vec k, m) = H_0(\vec k, m) + M \cos\theta \sigma_1 + M \sin\theta \sigma_2,
\end{align}
where we have assumed a uniform mass parameter.

\begin{figure}[!t]
\centering
\includegraphics[width=0.8\linewidth]{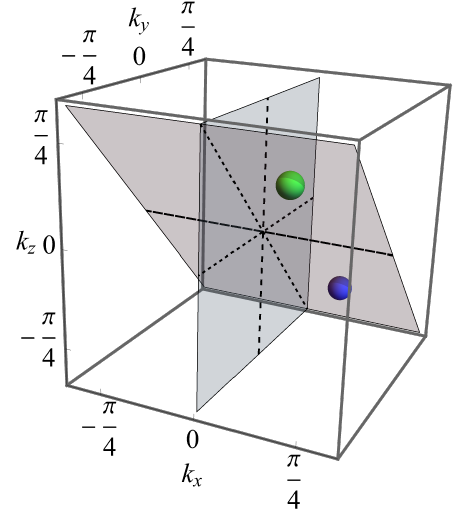}
\caption{Location of the Weyl points (colored spheres) in the interaction-driven Weyl semimetal phase. The asymmetric location is a result of a lack of inversion and mirror symmetries. The $k_x = 0$ and $k_y + k_z = 0$ planes are highlighted as a guide for the eye.}
\label{fig:WPs}
\end{figure}

Interaction-driven spontaneous rotational symmetry breaking at the TQCP ($m=3$) splits the QBT point into a pair of linear band crossings which carry unit Berry monopole charges, i.e.~a pair of Weyl points.  
While parameter $M$ controls the separation between the Weyl points, the angle $\theta$ controls their angular orientation relative to the principal axes of the Brillouin zone.
Away from the TQCP ($m\neq 3$), the WSM phase is separated from the Hopf insulator  ($m< 3$) and the trivial insulator ($m> 3$) by distinct symmetry-broken insulating states, as depicted in Fig.~\ref{fig:phase-diag}. 
The quantum phase transitions between the WSM phase and the symmetry-broken insulators are  tuned by $m$, and they are  characterized by the merging of the Weyl points. 
Thus, the entire phase boundary separating the WSM phase from the symmetry-broken insulators represents a \textit{line} of TQCPs, achieved by an interplay between the interaction strength, $g$, and the band-inversion parameter, $m$.
At all such TQCPs the bands disperse quadratically about the touching point. 

\subsection{Topology and surface states}\label{sec:topo}
While it is relatively straightforward to characterize the topology of the WSM phase, the topology of the symmetry-broken insulators is more subtle. 
Therefore, we begin with an analysis of the WSM phase, and then we will discuss the topological features in the symmetry-broken insulators.
Without loss of generality, in this subsection we will fix $\theta = 0$ in Eq.~\eqref{eq:MeanFieldLattice}.

\begin{figure}[!t]
\centering
\subfloat[]{%
\includegraphics[width=0.95\linewidth]{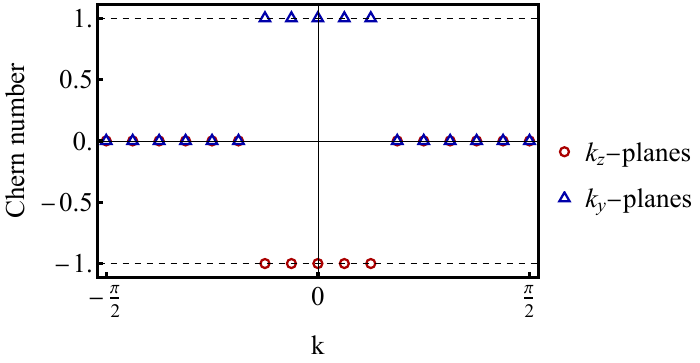}%
}
\hfill
\subfloat[]{%
\includegraphics[width=0.75\linewidth]{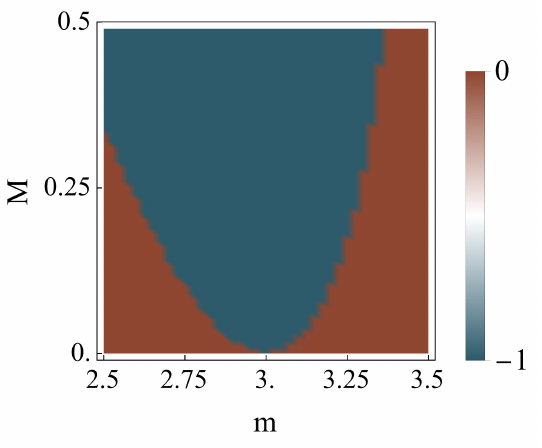}%
}
\caption{Chern number carried by planes separating the Weyl points. 
(a) Chern number per plane. 
(b) Chern number carried by the $k_z = 0$ plane as a function of the band-inversion parameter  $m$ and magnitude of the mass term, M. As shown in (a), it equals $-1$ in the Weyl semimetal phase and $0$ outside.}
\label{fig:chern-num}
\end{figure}

\begin{figure}
\centering
\includegraphics[width=1\linewidth]{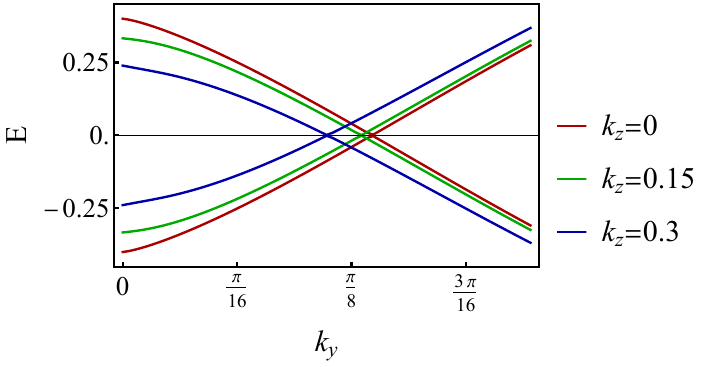}
\caption{States localized on the surface-terminations perpendicular to $\hat x$ in the slab-geometry. Each $k_z$-cut of the surface Brillouin zone represents the edge states supported by the corresponding $k_x - k_y$ plane in the bulk [cf. Figs.~\ref{fig:WPs} and \ref{fig:chern-num}(a)]. At each cut the two counter-propagating states exist on opposite sides of the slab. Taken together, the zero modes describe a Fermi arc in the surface Brillouin zone that does not lie along a high-symmetry axis, and curves away from the $k_z$-axis.}
\label{fig:arcs}
\end{figure}

At $\theta = 0$, the Weyl points are located on the plane spanned by the $[100]$ and $[01\bar{1}]$ axes of the Brillouin zone, as shown in Fig.~\ref{fig:WPs}.
The two-dimensional planes separating the Weyl points carry a non-trivial Chern number because the Weyl points act as  Berry curvature monopoles.
From Fig.~\ref{fig:WPs} we note that there are two sets of such planes, viz., $k_x-k_y$ planes and $k_x - k_z$ planes.
In Fig.~\ref{fig:chern-num}(a) we depict the Chern number supported by these planes.
The Chern number jumps across the planes that host a Weyl point. 
The presence of the Chern number on the $k_z = 0$ or $k_y = 0$ planes can be used to determine the phase boundary between the WSM phase and the symmetry-broken insulators, as shown in Fig.~\ref{fig:chern-num}(b).
Each plane supporting a finite Chern number can be viewed as a two-dimensional Chern insulator embedded within a three-dimensional Brillouin zone, such that the chiral edge states they support contribute to the topologically protected surface states of the three-dimensional system. 
Since in the WSM phase Chern insulating planes are layered along both $\hat k_z$ and $\hat k_y$, all standard surface terminations will host Fermi arcs, as exemplified by the $(100)$ surface in Fig.~\ref{fig:arcs}.

The symmetry-broken insulators do not enjoy the same degree of robust bulk topology and bulk-boundary correspondence as the WSM phase. 
Their bulk topology, however, can be diagnosed with the help of the RTP invariant.
In the Hopf insulating phase, the one-dimensional polarization,
\begin{align}
P_z(k_x, k_y) = \frac{1}{\pi}\int_{-\pi}^\pi \dd{k_z} A_z(\bs k) 
\end{align} 
with $A_z$ being the $z$-th component of the Berry connection associated with the conduction band, winds and unwinds from $0 \to 1 \to 0$ along both $X - \Gamma$ and $M - \Gamma$ directions in the $k_x - k_y$ plane of the Brillouin zone.
This non-trivial winding along two non-parallel and symmetry-inequivalent directions encode the non-triviality of the RTP invariant characterizing the Hopf insulator phase~\cite{nelson2022}.
By contrast, in the ``Nematic insulator \#1''  only the $X - \Gamma$ path supports a full winding of $P_z$ with $P_z$ winding  from $0 \to 1$ along a half-cycle of the path $M - \Gamma$.
$P_z$ does not wind along any path in the two insulating phases on the $m > 3$ side of the phase diagram, Fig.~\ref{fig:phase-diag}.
Therefore, we conclude that the symmetry-broken insulator on the $m<3$ side inherits a lower-dimensional topological features of the Hopf insulator. 
Since no robust bulk-boundary correspondence exists for the RTP invariant, we do not discuss the surface states in the symmetry-broken insulators as these are not expected to be topologically protected.

\section{Non-reciprocal superconductivity}\label{sec:SC}
The energy bands obtained by diagonalizing $H_{\text{MF}}(\vec k,m)$, $\varepsilon_\pm^{\text{MF}}(\vec k)$, are asymmetric under inversion, $\varepsilon_s^{\text{MF}}(-\vec k) \neq \varepsilon_s^{\text{MF}}(\vec k)$. 
The degree of asymmetry is controlled by the magnitude of the rotational-symmetry-breaking mass term, $[\varepsilon_s^{\text{MF}}(\vec k)]^2 - [\varepsilon_s^{\text{MF}}(-\vec k)]^2 \propto M$.
When these mean-field states are doped, the asymmetry implies that the group velocity of the quasi-particles on opposite sides of the Fermi surface are unrelated.
This is a sufficient condition for realizing a \textit{non-reciprocal metal}~\cite{davydova2024}.
Therefore, any superconducting state obtained by pairing states on the Fermi surface will be a non-reciprocal superconductor, supporting a superconducting diode effect~\cite{Nadeem2023,davydova2024}.
Assuming the existence of sub-dominant attractive interactions and a doping scale that is sufficiently weaker than the repulsive interaction considered above, in this section we will briefly discuss the properties of an orbital-singlet superconducting state without rotational symmetry. 

\begin{figure}
\includegraphics[width=0.9\columnwidth]{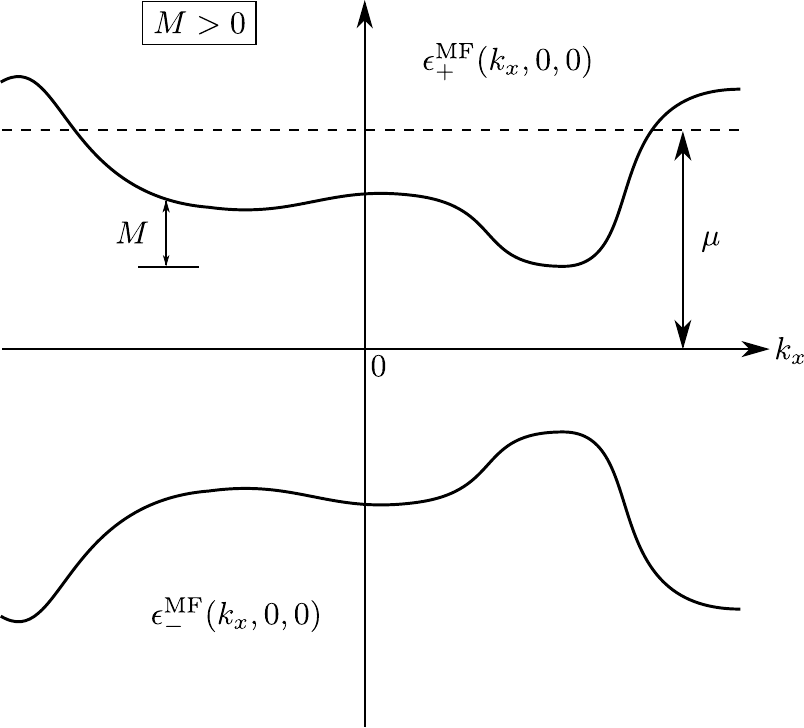}
\caption{Appearance of a single  Fermi pocket around the centre of the Brillouin zone in the presence of a rotational-symmetry-breaking mass term $M>0$ in Eq.~\eqref{eq:MeanFieldLattice} when $m=3$ (TQCP). For simplicity, the two electronic bands $\epsilon^\text{MF}_\pm(\vec k)$ are plotted along the $\hat{k}_x$ axis. The chemical potential $\mu$ is taken sufficiently large for a single Fermi pocket to be formed.}
\label{fig:NonReciprocalFermiPocket}
\end{figure}

\begin{figure}[!t]
\centering
\includegraphics[width=0.9\columnwidth]{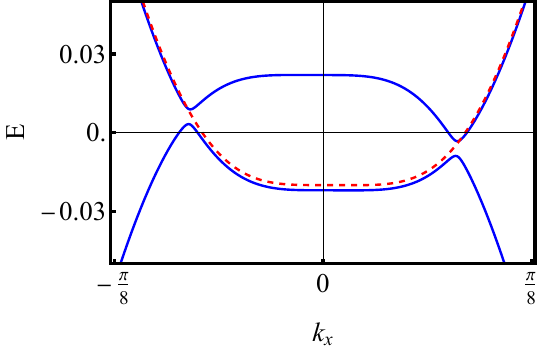}
\caption{A section of the spectrum obtained by diagonalizing the BdG Hamiltonian in Eq.~\ref{eq:BdG}. 
Although the spectrum contains four bands,  we have zoomed to the vicinity of zero energy to demonstrate the asymmetry in the spectrum and the presence of a Bogoliubov-Fermi surface. Here, the solid (dashed) curves indicate the BdG (normal state) bands.}
\label{fig:BdG}
\end{figure}

Owing to a lack of inversion,  mirror, and time-reversal   symmetries, generically the Fermi pockets obtained at finite dopings are not centered about a high-symmetry point in the Brillouin zone. 
Therefore, for simplicity, we primarily focus in the vicinity of $m = 3$ with a sufficiently large chemical potential such that a single Fermi pocket is present around the center of the Brillouin zone (see Fig.~\ref{fig:NonReciprocalFermiPocket}).
This configuration allows us to consider pairing channels carrying zero center-of-mass momentum. 
The Bogoliubov-de-Gennes (BdG) Hamiltonian matrix for an orbital-singlet superconductor is given by 
\begin{align}
H_{\text{BdG}}(\bs k, m) = \mqty[H_0(\vec k, m) - \mu &  \Delta_{\text{sc}} \sigma_2 \\  \Delta_{\text{sc}} \sigma_2  & \mu -H_0^*(-\vec k, m)],
\label{eq:BdG}
\end{align}
where the pairing gap $\Delta_{\text{sc}}$ is assumed to be a real constant. 
The BdG spectrum exhibits a pronounced asymmetry for $M \neq 0$, as exemplified by  Fig.~\ref{fig:BdG}. 
We note the presence of a two-dimensional manifold of zero energy states, i.e.~a Bogoliubov-Fermi surface~\cite{agterberg2017,Brydon2018,Link2020,Zhu2021b}. 
The volume enclosed by the Bogoliubov-Fermi surface is controlled by a competition between the nematic mass term, $M$, and the size of the superconducting gap, $\Delta_{sc}$. 
In particular, the condition $M > \Delta_{\text{sc}}$ is necessary for the existence of a Bogoliubov-Fermi surface.

Both the non-reciprocal superconductivity and the existence of a Bogoliubov-Fermi surface are general properties of all superconducting states resulting from pairing instabilities of the non-reciprocal metals obtained by doping any of the three rotational-symmetry broken states in Fig.~\ref{fig:phase-diag}.
However, away from the aforementioned parameter regime that leads to a single Fermi pocket located at the center of the Brillouin zone, the candidate pairing states  carry finite momentum due to the large anisotropy of the Fermi pockets. 

\section{Conclusion}\label{sec:conclusion}
In this paper, we have studied the effects of strong repulsive interaction on  four-fold rotationally symmetric Hopf insulators, and we showed that it predominantly drives a nematic instability,  characterized by a breaking of the rotational symmetry.
By anchoring our analysis to the TQCP  separating a Hopf insulator from a trivial insulator, we find three interaction-driven nematic phases: a nematic Weyl semimetal and two nematic insulators.
By a combination of an $\epsilon$-expansion scheme and mean-field theory, we obtained a global phase diagram of interacting Hopf insulators. 
We also discuss the topological features of the nematic states and the nature of the surface states in the WSM phase.

Our $\epsilon$-expansion scheme amounts to dimensional tuning, enabling us to connect the interacting TQCP in the model of the Hopf insulator to two-dimensional interacting quadratic band-touching semimetals. 
In particular, the anomalous Hall insulator, which nucleates as the dominant instability of the latter, is embedded within the WSM state obtained by destabilizing the TQCP as a Chern-number-carrying plane in the Brillouin zone. 
Thus, this anomalous Hall insulating plane represents one of the Chern insulating layers in the  Brillouin zone that separates the two Weyl points in the WSM phase.
We further note that the quadratic band-touching point at the TQCP is an example of an emergent Berry dipole, and it may be considered as a two-band variant of the quadratic band touching point in Luttinger semimetals. 
Since the latter also realizes nematic states in the presence of strong repulsive interactions~\cite{Goswami2017}, it would be of future interest to investigate what aspect of the correlation physics discussed here survives in correlated Luttinger semimetals, especially in connection to the influence of Berry dipoles on  correlation driven phenomena.

Finally, we demonstrated that if the nematic states undergo a superconducting transition due to residual attractive interactions or proximity induced pairing, then these pairing states are candidate non-reciprocal superconductors. 
Such superconducting state are of topical interest for their ability to sustain diode effects. 
Thus, our work identifies phases of matter with a propensity for supporting intrinsic superconducting diode effects. 
An additional but important feature that was not discussed in this context is that the superconductors obtained through a pairing instability of lightly doped WSM necessarily carry a finite vorticity~\cite{li2018topological}.
This is a manifestation of the existence of Berry monopoles (i.e. Weyl points) that are submerged within the Fermi pockets in the parent state.
The non-reciprocal superconductors obtained as instabilities of the WSM phase will also possess similar topology-enforced vorticity.
This presents us with a novel context for studying the role of monopole-obstruction in non-reciprocal superconductivity and the impact of monopole-obstructions in diode response.
We leave such fascinating considerations to future works.

\acknowledgements K.L. partially acknowledges funding by the Deutsche
Forschungsgemeinschaft (DFG) via the project A04 of the Collaborative Research
Center SFB 1143 (project-id 247310070), and the Cluster of
Excellence on Complexity and Topology in Quantum Matter
ct.qmat (EXC 2147, project-id 390858490). S.S. is supported by NSF (DMR-2220603) and AFOSR (FA9550-21-1-0356).


\renewcommand{\emph}{\textit}
\bibliography{hopf}

\clearpage


\appendix

\onecolumngrid 

\section{Renormalization Group}\label{sec:RenormalizationGroup}
Here we provide details regarding the derivation of the RG equations given in Eqs.~\eqref{eq:TopologicalMassRGEquation}, \eqref{eq:BetaFunction}, and \eqref{eq:SusceptibilitiesRGEquations} in the main text. To that end, we first extend the spatially three-dimensional action given in Eq.~\eqref{eq:action} to one in the \textit{fractional} dimension $d=2+\epsilon$ $(0\leq\epsilon\leq1)$ including a source term as well:
\begin{flalign}\label{eq:action_d}
	S_\text{<, one-loop}(\delta_b)=& T\sum_{i\omega_n}\int\dfrac{d^3k}{(2\pi)^d}\Psi^\dag(\vec{k},i\omega_n)\qty[i\omega_n\mathbbm{1}_2 - h_0(\vec{k}, \delta_b)]\Psi(\vec{k},i\omega_n)\nn\\
	&+\dfrac{g}{2}T\sum_{i\omega_n}\int\dfrac{d^dk}{(2\pi)^d}n_A(-\vec{k},-i\omega_n)n_B(\vec{k},i\omega_n)\nn\\
	&-\sum^3_{j=0}M_j\Psi^\dag(\vec{k},i\omega_n)\sigma_j\Psi(\vec{k},i\omega_n).
\end{flalign}
The vector $\vec{k}=(k_1,k_2,\dots,k_d)$ represents the $d$-dimensional momentum, and $\omega_n=(2n+1)\pi\,T$ stands for the fermionic Matsubara frequencies with $n$ being an integer-valued number. The single-particle Hamiltonian matrix $h_{0,d}(\vec{k}, \delta_b)$ reads
\begin{align}\label{eq:hamiltonian_d}
	 h_{0,d}(\vec{k}, \delta_b)=\sum^{d-1}_{j=1}2\qty[t^2k_jk_d+t\sum^{d-1}_{i=j+1}k_i\delta_b-t\sum^{j-1}_{l=1}k_l\delta_b]\gamma_j + \qty[t^2\left(\sum^{d-1}_{j=1}k^2_j-k^2_d\right)- \delta_b^2]\gamma_d.
\end{align}
The set $\{\gamma_1,\dots,\gamma_d\}$ consists of mutually anticommuting two-dimensional matrices defined through the relation $\{\gamma_n,\gamma_m\}\sim\delta_{nm}\mathbbm{1}_2$ for $n,m\in\{1,\dots,d\}$. Since the coupling constant is marginally relevant at $d=2$ spatial dimensions, we employ the one-loop Wilsonian RG scheme using the expansion parameter $\epsilon=d-2$ \cite{herbut2007,goldenfeld2018}. Such renormalization procedure is facilitated using the radial component of the momentum vector to define the successive RG momentum shells. In particular, the interval $\Lambda/\ell<k<\Lambda$ ($0<k<\Lambda/\ell$) defines the ``fast modes'' (``slow modes''), where $\Lambda$ is the momentum UV cutoff. The dimensionless scaling parameter $\ell$ takes real values within the interval $(1,\infty)$. The angular part of the involved Feynman diagrams is calculated at the fixed spatial dimension $d=3$. Subsequently, one can decompose the fermionic spinor shown in Eq.~\eqref{eq:action_d} into two parts associated to particular momentum shells
\begin{flalign}
	\Psi(\vec{k},i\omega_n)= \left\{
	\begin{array}{ll}
		\Psi_>(\vec{k},i\omega_n), &\quad \Lambda/\ell<k<\Lambda \\
		\Psi_<(\vec{k},i\omega_n), &\quad 0<k<\Lambda/\ell. \\
	\end{array} 
	\right. 
\end{flalign}
By integrating out the ``fast modes'' perturbatively to one-loop order, the corresponding \textit{renormalized} action for the ``slow modes'' reads
\begin{align}\label{eq:action_d_oneloop}
	S_\text{<, one-loop}(\delta_b)=& T\sum_{i\omega_n}\int\dfrac{d^dk}{(2\pi)^d}\Psi^\dag_<(\vec{k},i\omega_n)\qty[i\omega_n\mathbbm{1}_2 - h_{0,d}(\vec{k}, \delta_b)+\Sigma_{HF}(g, \delta_b)]\Psi_<(\vec{k},i\omega_n)\nn\\
	&+\left[\dfrac{g}{2}-V_g(g,\delta_b)\right]T\sum_{i\omega_n}\int\dfrac{d^dk}{(2\pi)^d}n_{<,A}(-\vec{k},-i\omega_n)n_{<,B}(\vec{k},i\omega_n)\nn\\
	&-\sum^d_{j=0}\Big[M_j+V^{(j)}_M(g,\delta_b,M_1,\dots,M_d)\Big]\Psi^\dag_<(\vec{k},i\omega_n)\gamma_j\Psi_<(\vec{k},i\omega_n)
\end{align}
The term $\Sigma_{HF}(g, \delta_b)$ stands for the one-loop fermionic self energy, while the terms $V_g(g,\delta_b)$ and $V^{(j)}_M(g,\delta_b,M_1,\dots,M_d)$ represent the one-loop vertex corrections to the four-fermion interaction and source term, respectively.

\subsection{Topological mass renormalization}
To focus on the renormalization of the bare topological mass, we set $V_g(g,\delta_b)=V^{(j)}_M(g,\delta_b,M_1,\dots,M_d)=0$ in Eq.~\eqref{eq:action_d_oneloop}:
\begin{align}\label{eq:action_d_oneloop_1}
	S^{(1)}_\text{<, one-loop}(\delta_b)=& T\sum_{i\omega_n}\int\dfrac{d^dk}{(2\pi)^d}\Psi^\dag_<(\vec{k},i\omega_n)\qty[i\omega_n\mathbbm{1}_2 - h_{0,d}(\vec{k}, \delta_b)+\Sigma_{HF}(g, \delta_b)]\Psi_<(\vec{k},i\omega_n)\nn\\
	&+\dfrac{g}{2}\,T\sum_{i\omega_n}\int\dfrac{d^dk}{(2\pi)^d}n_{<,A}(-\vec{k},-i\omega_n)n_{<,B}(\vec{k},i\omega_n)\nn\\
	&-\sum^d_{j=0}M_j\Psi^\dag_<(\vec{k},i\omega_n)\gamma_j\Psi_<(\vec{k},i\omega_n).
\end{align}
Denoting the $d$-dimensional single-particle Matsubara Green's function for the ``fast modes'' as $G_>(\vec{q},i\nu_n)$, with $d$-dimensional momentum vector $\vec q$ and fermionic Matsubara frequency $\nu_n$, the self energy $\Sigma_{HF}(g, \delta_b)$ writes
\begin{figure}
    \centering
    \subfloat[]{\includegraphics[width=0.15\linewidth]{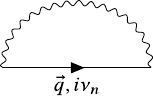}}\hspace*{2cm}
	\subfloat[]{\includegraphics[width=0.08\linewidth]{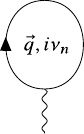}}
	\label{fig:hartree-fock}
	\caption{Self-energy contributions from one-loop diagrams, (a) Hartree diagram, and (b) Fock diagram. The continuous line with an arrow denotes the $d$-dimensional single-particle Matsubara Green's function for the ``fast modes'' $G_>(\vec{q},i\nu_n)$.}
\end{figure}
\begin{flalign}
	\Sigma_{HF}(g, \delta_b)=g\,T\sum_{i\nu_n}\int\dfrac{d^dq}{(2\pi)^d}\bigg(-\text{Tr}\Big[G_>(\vec{q},i\nu_n)\Big]+G_>(\vec{q},i\nu_n)\bigg).
\end{flalign}
Pictorially, it is given in terms of the Hartree-Fock diagrammatic contributions shown in Fig.~\ref{fig:hartree-fock} \cite{fetterwalecka}. Summing over the Matsubara frequencies and performing the momentum integration in the limit $\epsilon\to0$ at zero temperature $(T\to0)$, we arrive at the expression
\begin{flalign}\label{eq:selfenergy_d_1}
	 \Sigma_{HF}(g, \delta_b)=g\,\ell^2\left[\dfrac{\Lambda^{2+\epsilon}(1-\ell^{-2})}{24\pi^2}-\dfrac{\Lambda^{\epsilon}\,\text{ln}\ell\,\delta^2_b}{3\pi^2t^2}\right]\gamma_d.
\end{flalign}
Eq.~\eqref{eq:selfenergy_d_1} leads to a nontrivial renormalization for the squared topological mass term $\delta^2_b$, which is part of the \textit{form factor} multiplying the matrix $\gamma_d$. Upon introducing the following dimensionless quantities
\begin{flalign}\label{eq:dimless}
	&\widetilde{g}:=\dfrac{g}{8\pi^2}\left(\dfrac{\Lambda}{\ell}\right)^{\epsilon},\qquad	\widetilde{\Psi}:=\Psi\left(\dfrac{\Lambda}{\ell}\right)^{-(4+d)/2},\qquad
	\widetilde{k}:=k\left(\dfrac{\Lambda}{\ell}\right)^{-1},\qquad\widetilde{\omega}_n:=\omega_n\left(\dfrac{\Lambda}{\ell}\right)^{-2},\nn\\
	&\widetilde{T}:=T\left(\dfrac{\Lambda}{\ell}\right)^{-2},\qquad\widetilde{\delta_b}:=\delta_b\left(\dfrac{\Lambda}{\ell}\right)^{-1},\quad \widetilde{M}_j:=M_j\left(\dfrac{\Lambda}{\ell}\right)^{-2},
\end{flalign}
the action in Eq.~\eqref{eq:action_d_oneloop_1} becomes \textit{self-similar} to the corresponding part given in Eq.~\eqref{eq:action_d}, but with renormalized squared topological mass term determined by the RG recursion formula
\begin{equation}\label{eq:recursion_formula_delta}
	\widetilde{\delta}^2_\ell:=\widetilde{\delta}^2_b\,\ell^2-\widetilde{g}\,\ell^2\left[\dfrac{1-\ell^{-2}}{24\pi^2}-\dfrac{8\,\text{ln}\ell\,\widetilde{\delta}^2_b}{3t^2}\right].
\end{equation}
The first term inside the square brackets in Eq.~\eqref{eq:recursion_formula_delta} has a non-universal dependence on the step used for the applied scale transformations, and it shifts the zero-mass location for the squared bare topological mass from zero to the value $\Delta_\text{TQCP}=\widetilde{g}(1-\ell^{-2})/(24\pi^2)$ \cite{herbut2007,goldenfeld2018}. We, thus, define the \textit{difference} between the squared bare topological mass and the zero-mass location to be
\begin{equation}
	\widetilde{\Delta}:=\widetilde{\delta}^2_b-\widetilde{\Delta}_\text{TQCP}.
\end{equation}
The associated RG recursion formula then reads
\begin{equation}
\widetilde{\Delta}_\ell=\widetilde{\Delta}\,\ell^2\left[1+\dfrac{8\,\widetilde{g}\,\text{ln}\ell}{3t^2}\right].
\end{equation}
The corresponding RG equation for the dimensionless quantity $\widetilde{\Delta}$ is found by considering an infinitesimal scale transformation $\ell\to\ell+d\ell$:
\begin{equation}
\dfrac{d\widetilde{\Delta}}{d\text{ln}\ell}=\widetilde{\Delta}\left[2+\dfrac{8\,\widetilde{g}}{3t^2}\right].
\label{eq:rg_equation_Delta}
\end{equation}
Eq.~\eqref{eq:rg_equation_Delta} coincides with Eq.~\eqref{eq:TopologicalMassRGEquation} from the main text.

\subsection{Coupling constant renormalization}
Setting $\Sigma_{HF}(g, \delta_b)=V^{(j)}_M(g,\delta_b,M_1,\dots,M_d)=0$ in Eq.~\eqref{eq:action_d_oneloop}, the $d$-dimensional action for the renormalization of the interaction vertex writes
\begin{figure}
    \centering
    \subfloat[]{\includegraphics[width=0.075\linewidth]{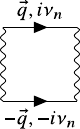}}\hspace*{2cm}
    \subfloat[]{\includegraphics[width=0.07\linewidth]{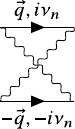}}\hspace*{2cm}
    \subfloat[]{\includegraphics[width=0.15\linewidth]{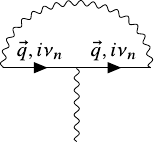}}\hspace*{2cm}
    \subfloat[]{\includegraphics[width=0.075\linewidth]{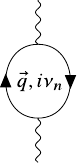}}
    \caption{Diagrammatic contributions to the vertex correction given in Eq.~\eqref{eq:vertexcorrection_d_1}.}
	\label{fig:vertexcorrections}
\end{figure}
\begin{align}\label{eq:action_d_oneloop_2}
	S^{(2)}_\text{<, one-loop}(\delta_b)=& T\sum_{i\omega_n}\int\dfrac{d^dk}{(2\pi)^d}\Psi^\dag_<(\vec{k},i\omega_n)\qty[i\omega_n\mathbbm{1}_2 - h_{0,d}(\vec{k}, \delta_b)]\Psi_<(\vec{k},i\omega_n)\nn\\
	&+\left[\dfrac{g}{2}-V(g,\delta_b)\right]\,T\sum_{i\omega_n}\int\dfrac{d^dk}{(2\pi)^d}n_{<,A}(-\vec{k},-i\omega_n)n_{<,B}(\vec{k},i\omega_n)\nn\\
	&-\sum^d_{j=0}M_j\Psi^\dag_<(\vec{k},i\omega_n)\gamma_j\Psi_<(\vec{k},i\omega_n)
\end{align}
In terms of the $d$-dimensional single-particle Matsubara Green's function for the ``fast modes'' $G_>(\vec{q},i\nu_n)$, the one-loop vertex correction $V(g, \delta_b)$ reads
\begin{flalign}\label{eq:vertexcorrection_d_1}
		V(g,\delta_b)=
		\dfrac{g^2}{2}\,T\sum_{i\nu_n}\int\dfrac{d^dq}{(2\pi)^d}\bigg(&-2\,\text{Tr}\Big[G_>(\vec{q},i\nu_n)G_>(-\vec{q},-i\nu_n)\Big]+2\,\text{Tr}\Big[G_>(\vec{q},i\nu_n)\Big]\text{Tr}\Big[G_>(-\vec{q},-i\nu_n)\Big]\nn\\
		&+6\,\text{Tr}\Big[G_>(\vec{q},i\nu_n)G_>(\vec{q},i\nu_n)\Big]+2\,\text{Tr}\Big[G_>(\vec{q},i\nu_n)\Big]\text{Tr}\Big[G_>(\vec{q},i\nu_n)\Big]\bigg).
\end{flalign}
The corresponding diagrammatic contributions are pictorially shown in Fig.~\ref{fig:vertexcorrections} \cite{fetterwalecka}. Summing over the Matsubara frequencies and integrating over the momentum values in the limit $\epsilon\to0$ at zero temperature, Eq.~\eqref{eq:vertexcorrection_d_1} becomes 
\begin{equation}
	V(g,\delta_b)=\dfrac{g^2\,\Lambda^\epsilon\,\text{ln}\ell}{8\pi^2t^2}.
\end{equation}
Upon inserted into Eq.~\eqref{eq:action_d_oneloop_2}, the latter result leads to a scale-dependent renormalization of the coupling constant. Furthermore, we introduce the dimensionless quantities given in Eq.~\eqref{eq:dimless} such that the aforementioned action becomes dimensionless. The renormalized coupling constant then obeys the following RG recursion formula
\begin{equation}\label{eq:couplingconstant_recursionformula}
	\widetilde{g}_\ell=\ell^{-\epsilon}\left[\widetilde{g}-\widetilde{g}^2\,\text{ln}\ell\,\dfrac{2}{t^2}\right].
\end{equation}
Considering an infinitesimal scale transformation $\ell\to\ell+d\ell$ in Eq.~\eqref{eq:couplingconstant_recursionformula}, we arrive at the RG equation for the dimensionless coupling constant
\begin{equation}
\dfrac{d\widetilde{g}}{d\text{ln}\ell}=-\epsilon\,\widetilde{g} +\dfrac{2\widetilde{g}^2}{t^2}.
\end{equation}

\subsection{Source term renormalization}
Finally, setting  $\Sigma_{HF}(g, \delta_b)=V_g(g,\delta_b)=0$ in Eq.~\eqref{eq:action_d_oneloop}, we obtain the $d$-dimensional action for the renormalization of the source term:
\begin{align}
	S^{(3)}_\text{<, one-loop}(\delta_b)=& T\sum_{i\omega_n}\int\dfrac{d^dk}{(2\pi)^d}\Psi^\dag_<(\vec{k},i\omega_n)\qty[i\omega_n\mathbbm{1}_2 - h_{0,d}(\vec{k}, \delta_b)]\Psi_<(\vec{k},i\omega_n)\nn\\
	&+\dfrac{g}{2}T\sum_{i\omega_n}\int\dfrac{d^dk}{(2\pi)^d}n_{<,A}(-\vec{k},-i\omega_n)n_{<,B}(\vec{k},i\omega_n)\nn\\
	&-\sum^d_{j=0}\Big[M_j+V^{(j)}_M(g,\delta_b,M_1,\dots,M_d)\Big]\Psi^\dag_<(\vec{k},i\omega_n)\gamma_j\Psi_<(\vec{k},i\omega_n).
\end{align}
The vertex correction $V^{(j)}_M(g,\delta_b,M_1,\dots,M_d)$ reads
\begin{figure}
    \centering
    \subfloat[]{\includegraphics[width=0.15\linewidth]{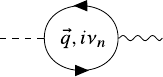}}\hspace*{2cm}
    \subfloat[]{\includegraphics[width=0.1\linewidth]{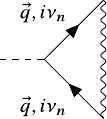}}
    \caption{Vertex corrections to the source term given in Eq.~\eqref{eq:vertexcorrection_d_2}. The dashed line denotes the matrix $\gamma_j$ corresponding to a particular mass term.}
\label{fig:vertexcorrections_source}
\end{figure}
\begin{flalign}\label{eq:vertexcorrection_d_2}
	V^{(j)}_M(g,\delta_b,M_1,\dots,M_d)=gM_jT\sum_{i\nu_n}\int\dfrac{d^dq}{(2\pi)^d}\bigg(\text{Tr}\Big[G_>(\vec{q},i\nu_n)\gamma_jG_>(\vec{q},i\nu_n)\gamma_j\Big]-\text{Tr}\Big[G_>(\vec{q},i\nu_n)\gamma_jG_>(\vec{q},i\nu_n)\Big]\text{Tr}\Big[\gamma_j\Big]\bigg),
\end{flalign}
and the corresponding one-loop diagrammatic contributions are shown in Fig.~\ref{fig:vertexcorrections_source}. Considering the case of a source term in three spatial dimensions $(j=0,1,2,3)$, the latter expression yields in the limit $\epsilon\to0$ at zero temperature
\begin{flalign}
	V^{(j)}_M(g,\delta_b,M_1,\dots,M_d)=gM_j\dfrac{\Lambda^\epsilon\text{ln}\ell}{8\pi^2t^2}\left\{
	\begin{array}{ll}
		0,\qquad j=0,\\\nn\\
		\dfrac{22}{15},\qquad j=1,2,\\\nn\\
		\dfrac{16}{15},\qquad j=3.
	\end{array} 
	\right.
\end{flalign}
Inserting the latter result into Eq.~\eqref{eq:action_d_oneloop_2} and using the dimensionless quantities given in Eq.~\eqref{eq:dimless}, we derive the RG recursion expression for the $j$-th dimensionless mass parameter $\widetilde{M}_j$,
\begin{equation}\label{eq:sourceterm_recursionformula}
	\widetilde{M}_{j,\ell}=\ell^2\bigg[\widetilde{M}_j+V^{(j)}_M(g,\delta_b,M_1,\dots,M_d)\bigg].
\end{equation}
Upon performing the infinitesimal scale transformation $\ell\to\ell+d\ell$ in Eq.~\eqref{eq:sourceterm_recursionformula}, we arrive at the following RG equation
\begin{equation}\label{eq:sourceterm_RG}
	\dfrac{d\text{ln}\widetilde{M}_j}{d\text{ln}\ell}=2+\dfrac{\widetilde{g}}{t^2}\left\{
	\begin{array}{ll}
		0,\qquad j=0\\\\
		\dfrac{22}{15},\qquad j=1,2\\\\
		\dfrac{16}{15},\qquad j=3.
	\end{array} 
	\right.
\end{equation}
Eq.~\eqref{eq:sourceterm_RG} coincides with Eq.~\eqref{eq:SusceptibilitiesRGEquations} from the main text.

\end{document}